\documentclass[pra,aps,twocolumn,superscriptaddress,twoside,showpacs]{revtex4-1}

\usepackage{amsmath,amsfonts,amssymb,amsthm,mathtools}

\usepackage[english]{babel}
\usepackage[utf8]{inputenc}

\usepackage{latexsym}
\usepackage[pdftex]{graphicx}
\usepackage{epstopdf}

\usepackage{fancyhdr}
\usepackage{relsize}

\usepackage{subfigure}
\usepackage{MnSymbol}

\usepackage{hyperref}

 \let\b=\beta      \let\d=\delta     
  \let\h=\eta           \let\l=\lambda
          \let\x=\xi     \let\p=\pi        \let\r=\rho
 \let\t=\tau      \let\ph=\varphi   
   \let\o=\omega  
\let\G=\Gamma

\def\vF{v_F}
\def\kF{k_F}
\def\Hl{H}
\def\vl{v}

\def\ol{\o}
\def\hl{\h}
\def\ul{u}

\def\io{\infty}
\def\eps{\epsilon}
\def\dw{\d}

\def\cHl{\mathcal{H}}
\def\cGl{\mathcal{G}}
\def\cJl{\mathcal{J}}
\def\hV{\hat{V}}

\def\nn{\nonumber}

\def\sgn{\operatorname{sgn}}
\def\Tr{\operatorname{Tr}}
\def\tr{\operatorname{tr}}

\def\Dx{\xi}
\def\Dt{\tau}

\newcommand{\ppr}{^{\vphantom{\prime}}}
\newcommand{\wick}[1]{\left. :\! \hspace{-0.5pt} #1 \hspace{-0.5pt} \!: \right.}

\newcommand{\R}{\mathbb{R}}

\newcommand{\cO}{\mathcal{O}}

\newcommand{\cW}{\mathcal{W}}
\newcommand{\cH}{\mathcal{H}}

\newcommand{\cC}{\mathcal{C}}

\newcommand{\cJ}{\mathcal{J}}
\newcommand{\cG}{\mathcal{G}}

\begin{document}

\title{Time evolution of the Luttinger model with nonuniform temperature profile}

\author{Edwin Langmann}
\email[Electronic address: ]{langmann@kth.se}
\affiliation{Department of Physics, KTH Royal Institute of Technology,
106 91 Stockholm, Sweden}

\author{Joel L.\ Lebowitz}
\email[Electronic address: ]{lebowitz@math.rutgers.edu}
\affiliation{Departments of Mathematics and Physics, Rutgers University, Piscataway,
New Jersey 08854, USA}
\affiliation{Institute for Advanced Study, Princeton, New Jersey 08540, USA}

\author{Vieri Mastropietro}
\email[Electronic address: ]{vieri.mastropietro@unimi.it}
\affiliation{Dipartimento di Matematica, Universit{\`a} degli Studi di Milano,
20133 Milano, Italy}

\author{Per Moosavi}
\email[Electronic address: ]{pmoosavi@kth.se}
\affiliation{Department of Physics, KTH Royal Institute of Technology,
106 91 Stockholm, Sweden}

\date{June 27, 2017}

%------------------------------------------

\begin{abstract}
We study the time evolution of a one-dimensional interacting fermion system described by the Luttinger model starting from a nonequilibrium state defined by a smooth temperature profile $T(x)$.
As a specific example we consider the case when $T(x)$ is equal to $T_L$ ($T_R$) far to the left (right).
Using a series expansion in $\epsilon = 2(T_{R} - T_{L})/(T_{L}+T_{R})$, we compute the energy density, the heat current density, and the fermion two-point correlation function for all times $t \geq 0$.
For local (delta-function) interactions, the first two are computed to all orders, giving simple exact expressions involving the Schwarzian derivative of the integral of $T(x)$.
For nonlocal interactions, breaking scale invariance, we compute the nonequilibrium steady state (NESS) to all orders and the evolution to first order in $\epsilon$.
The heat current in the NESS is universal even when conformal invariance is broken by the interactions, and its dependence on $T_{L,R}$ agrees with numerical results for the $XXZ$ spin chain.
Moreover, our analytical formulas predict peaks at short times in the transition region between different temperatures and show dispersion effects that, even if nonuniversal, are qualitatively similar to ones observed in numerical simulations for related models, such as spin chains and interacting lattice fermions.
\end{abstract}


%%%%%%%%%%%%%%%%%%%%%%%%%%%%%%%%%%%%%%%%%%%%%%%%%%%%%%%%%%%%%%%%%%%%%%%%%%%%%

\maketitle

%%%%%%%%%%%%%%%%%%%%%%%%%%%%%%%%%%%%%%%%%%%%%%%%%%%%%%%%%%%%%%%%%%%%%%%%%%%%%

\section{Introduction}
Experiments on ultracold atomic gases have led to renewed interest in nonequilibrium properties of isolated one-dimensional quantum systems \cite{EFG, BDZ, PSSV, RDO, LanEtAl, EsFa}.
This field also has roots in a rich history of theoretical works studying both classical \cite{RLL, SpLe, Dh, BLRb, LLP, BO, BOS} and quantum systems \cite{KaFi, Rue, JP1, JP2, AJPP, BJP, ARRS, LaMi, SaMi, SPA, PVC} out of equilibrium.
One often studied protocol is to join, at time $t = 0$, disconnected left and right parts of an infinite system, where each part is in thermal equilibrium with temperatures $T_L$ and $T_R$, respectively.
For $t>0$ the system is evolved with a fully translation invariant Hamiltonian; this produces a heat current and, for long times, the system tends to a {\it nonequilibrium steady state} (NESS) if $T_{L} \neq T_{R}$.
This is usually referred to as the {\it partitioning protocol}.

Using the above protocol, exact results for the NESS were obtained for simple integrable models such as the $XX$ and $XY$ spin chains using $C^{*}$-algebraic methods \cite{HoAr, AsPi, Og1, Og2} and nonequilibrium Green's functions (Keldysh formalism) \cite{ALA}.
When written in terms of fermions, these models are all {\it noninteracting}: they can be mapped to one-dimensional systems of spinless lattice fermions with Hamiltonians that are quadratic in the fermion fields.
For general systems of free lattice fermions, results for the NESS were obtained using a generalized Landauer-B{\"u}ttiker formula in \cite{AJPP, BJP}.

For {\it interacting} fermions the partitioning protocol was successfully used to obtain exact results for systems described by conformal field theories (CFTs) \cite{BeDo1, BeDo2, BeDo3, SoCa, HoLo}.
Beyond CFT there are otherwise few exact results for the NESS, and even fewer for the evolution, of interacting fermions; see, e.g., \cite{Ca, IuCa, RSM, KRSM, MaWa, LLMM, CaCh}.
Using the same protocol, the time evolution and properties of the NESS have been studied extensively numerically \cite{KIM, VKM, BLVRMF} and by approximate analytical methods \cite{LiAn, LVBD, LVMR} in various models.
Recently, effective hydrodynamic equations for the long-time and large-distance dynamics for Bethe ansatz-solvable models were proposed \cite{BCNF, CBT}; see also \cite{Bre, Spo}.
We also mention recent studies of the heat current and the thermal Drude weight based on Bethe ansatz \cite{Zot}, density matrix renormalization group \cite{Kar}, and hydrodynamics \cite{IlNa, BVKM1, BVKM2}.

Most results for systems of interacting fermions, such as those mentioned above, rely on approximate methods or on assumptions, and it is thus interesting to obtain exact results for specific models that can serve as a benchmark.
In this paper, we present some exact results for the {\it full time evolution} (not just the NESS) of a continuum system of interacting fermions described by the Luttinger model \cite{To, Th, Lu, MaLi} on the real line starting, at $t=0$, from a {\it nonequilibrium state} defined by a smooth temperature profile $T(x)$.
This is related to but different from the partitioning protocol.
Specifically, if $\cHl(x)$ is the energy density operator defining the Hamiltonian, $\Hl = \int dx \, \cHl(x)$, then the initial state is given by $\hat{\r} = e^{-\cGl} / \Tr e^{-\cGl}$, with
\begin{equation}
\label{dm_neq}
\cGl = \int dx \, \b(x)\cHl(x),
\end{equation}
where $\b(x) \equiv T(x)^{-1} = \b [1+\eps W(x)]$ for some smooth function $W(x)$, with $\b$ the average inverse temperature and $\eps$ the distance from equilibrium.
(We use units such that $\hbar = k_{B} = 1$.)
We will mainly be concerned with the case of a step-like profile $T(x)$ equal to $T_L$ ($T_R$) far to the left (right), e.g., $W(x) = -(1/2)\tanh(x/\dw)$ with $\dw>0$, where $\b$ and $\eps$ are determined by $\b(\mp \io) = T_{L,R}^{-1}$.
The evolution of the system is given by $H$, and we are interested in {\it nonequilibrium expectation values} ($\eps \neq 0$) of local observables $\cO$,
\begin{equation}
\label{cOt}
\langle \cO(t) \rangle
\equiv \Tr \hat{\r} \cO(t),
\end{equation}
where $\cO(t) = e^{i\Hl t} \cO e^{-i\Hl t}$. 
If $\eps = 0$, then $\langle \cO(t) \rangle = \langle \cO \rangle_{\b}$ is an equilibrium expectation value with temperature $T = \b^{-1}$.
For the Luttinger model, such equilibrium properties are well known, since a long time, from the celebrated exact solution in \cite{MaLi} using bosonization; see also, e.g., \cite{Hal, Gia, GNT, Kop, HSU, MEJ, LaMo, Voit}.

We use a series expansion in $\eps$ to compute the time evolution and the NESS for the Luttinger model both in the case of local (delta-function) and nonlocal interactions starting from a nonequilibrium state.
We show that the NESS is factorized in terms of the eigenmodes of the interacting Hamiltonian (plasmons) \cite{MaLi} and not in terms of the fermions;
the presence of interactions is manifested by interaction-dependent exponents in the fermion two-point correlation function.
In contrast, we find that the final heat current is universal even for interactions that break conformal invariance, and the form of its dependence on $T_{L,R}$ confirms previous numerical results for interacting lattice fermion models, such as the $XXZ$ spin chain studied in \cite{KIM}.
For noninteracting lattice models, such results for the temperature dependence were
obtained analytically in \cite{HoAr, AJPP, BJP, ALA}.

For local interactions, our series for the energy and heat current densities can be summed into exact formulas for the time evolution.
These results contain the {\it Schwarzian derivative} \cite{FMS} of the integral of $T(x)$, which is very suggestive in view of the conformal invariance in the local case.
Its presence produces peaks in the energy and heat current densities at zero time in the transition region between different temperatures.
These resemble what is found numerically in related models \cite{KIM, LVBD}, even if the shapes of such peaks clearly are nonuniversal. 

For nonlocal interactions, breaking conformal invariance, we obtain analytical results for the NESS to all orders and for the time evolution to first order in $\eps$.
In this case, dispersive effects appear in the evolution, which look qualitatively similar to those seen numerically in lattice models.
(Such dispersive effects are absent for local interactions.)

The following two methods are used to compute nonequilibrium expectation values: Method 1, based on the Dyson series, and Method 2, using one-particle operators; see Secs.~\ref{SubSec:Method1}~and~\ref{SubSec:Method2}, respectively. 
Method 1 allows one to compute nonequilibrium results to first order in $\eps$ from equilibrium ones, and it can be used even for non-exactly-solvable models.
Method 2 allows one to compute results for the Luttinger model to all orders in $\eps$, and it is in general applicable only to models that are quasi-free.

We consider the Luttinger model given by
\begin{gather}
\Hl = \sum_{r} \frac{1}{2} \int dx \,
			\left[
				\! \wick{ \psi^+_{r}(x) \left(-ir \vF \partial_x \right) \psi^-_{r}(x) }
				+ \textnormal{h.c.}
			\right] \\
+ \l \sum_{r,r'} \int dxdy \, V(x-y)
	\! \wick{ \psi^+_{r\ppr}(x) \psi^-_{r\ppr}(x) } \!
	\! \wick{ \psi^+_{r'}(y) \psi^-_{r'}(y) } \! \nn ,
\end{gather}
with fermion fields $\psi_r^{-}(x)$ and $\psi^{+}_{r}(x) = \psi^{-}_{r}(x)^{\dagger}$, where $r=+(-)$ denotes right- (left-) moving fermions, $\wick{\cdots}$ indicates Wick (normal) ordering, $\vF > 0$ is the Fermi velocity, $V(x)$ is the interaction potential, and $\l$ is the coupling constant.
We use notation similar to \cite{LLMM, MaLi}; cf.\ also \cite{MaWa, LaMo} and references therein.
Let $\hat V(p) = \int dx \, V(x) e^{-ipx}$ denote the Fourier transform of the potential.
The interactions must satisfy $\l \hV(p) > -\pi\vF/2$, and $V(x)$ can be local, $V(x) = \pi\vF \d(x)/2$, which requires renormalizations, or nonlocal with an interaction range $a>0$, e.g., $V(x) = \pi\vF/[4a\cosh(\pi x/2a)]$.
The above examples of potentials are used in Figs.~\ref{Fig:ed_0_at_t0_and_evolution_Luttinger}~and~\ref{Fig:jd_vs_t_Luttinger} below to illustrate our analytical results, but we emphasize that these results hold true for a large class of interactions \cite{LLMM, LaMo}.

In what follows, we study the evolution of the energy density $E(x,t) \equiv \langle \cHl(x,t) \rangle$, the heat current density $J(x,t) \equiv \langle \cJl(x,t) \rangle$, and the fermion two-point correlation function $S_{r}(\Dx,\Dt,x,t) \equiv \langle \psi^{+}_{r}(x+\Dx,t+\Dt)\psi^{-}_{r}(x,t) \rangle$, where $\cJl(x,t)$ is determined by the continuity equation $\partial_t \cHl(x,t) + \partial_x \cJl(x,t) = 0$. 
We start in Sec.~\ref{Sec:NESS} by presenting results for the NESS.
This serves as a useful benchmark for the finite-time results presented in Sec.~\ref{Sec:FT_results:Local_interactions} for local interactions and in Sec.~\ref{Sec:FT_results:Nonlocal_interactions} for nonlocal interactions.
Our methods are described in Sec.~\ref{Sec:Methods}, and concluding remarks are given in Sec.~\ref{Sec:Conclusions}. 
Some computational details are deferred to the Appendix. 

%%%%%%%%%%%%%%%%%%%%%%%%%%%%%%%%%%%%%%%%%%%%%%%%%%%%%%%%%%%%%%%%%%%%%%%%%%%%%

\section{Nonequilibrium steady state}
\label{Sec:NESS}
It is well known that the Fourier modes of the fermion densities, $\r_r(p) \equiv \int dx \, \wick{ \psi^+_{r\ppr}(x)\psi^-_{r\ppr}(x) } e^{-ipx}$, define boson operators \cite{MaLi}, and that the Luttinger Hamiltonian can be written as $\Hl = H_{+} + H_{-}$, with
\begin{equation}
\label{H_r}
H_{r} = \frac{1}{2} \int dq \, \vl(q) \! \wick{ \tilde{\r}_{r}(-q)\tilde{\r}_{r}(q) } \! ,
\end{equation}
using Bogoliubov transformed fermion densities $\tilde{\r}_r(p) = \r_{r}(p) \cosh\ph(p) - \r_{-r}(p) \sinh\ph(p)$, where $\tanh2\ph(p) = -\l\hV(p)/[\pi\vF + \l\hV(p)]$, and the renormalized Fermi velocity $\vl(p) = \vF \sqrt{1 + 2\l\hV(p)/\pi \vF}$ \cite{MaLi, LLMM, Voit}.
The $\tilde{\r}_r(p)$ are commonly referred to as {\em plasmons}, and the Luttinger Hamiltonian is diagonal in terms of these \cite{MaLi}.
To find the NESS, we write $\hat{\r}(t) = e^{-i\Hl t} \hat{\r} e^{i\Hl t} = e^{-\cGl(-t)}/\Tr (e^{-\cGl(-t)})$ with $\cGl(t) = \int dx \, \b(x) \cHl(x,t)$ and express $\cHl(x,t)$ in terms of $\tilde{\r}_r(p,t) = \tilde{\r}_r(p) e^{-ir\ol(p)t}$, where $\ol(p) \equiv \vl(p)p$.
Taking $t \to \io$ in $\hat{\r}(t)$ by making use of the Riemann-Lebesgue lemma (cf., e.g., \cite{LLMM}), which can be justified for expectation values using Method 2, we find
\begin{equation}
\label{NESS}
\lim_{t\to\io}
\Tr \hat{\r}(t) \cO
= \frac{ \Tr e^{-\b_{+}H_{+}-\b_{-}H_{-}} \cO}{ \Tr e^{-\b_{+}H_{+}-\b_{-}H_{-}} },
\end{equation}
with $\b_{+} = T_{L}^{-1}$ and $\b_{-} = T_{R}^{-1}$.
This NESS describes a translation invariant state factorized into right- and left-moving plasmons at equilibrium with temperatures $T_{\pm} = \b_{\pm}^{-1}$.
A similar NESS was obtained in \cite{BeDo1, BeDo2, BeDo3} for CFTs and in \cite{HoAr, AsPi, Og1, Og2} for the $XX$ chain; in the latter case the same factorization of the NESS is valid also in terms of right- and left-moving {\em fermions}, whereas in our case only the plasmons factorize in such a way but not the fermions.

The long-time limit of expectation values for all local observables can be computed using \eqref{NESS} by straightforward generalizations of well-known equilibrium computations.
By recalling that $\int dx\, \cHl(x) = \sum_{r} H_{r}$ with $H_{r}$ in \eqref{H_r} and using the continuity equation to show that $\int dx\, \cJl(x) = \sum_{r} ({r}/{2}) \int dq\, ({d\ol(q)}/{dq}) \vl(q) \! \wick{ \tilde{\r}_{r}(-q)\tilde{\r}_{r}(q) }$, we obtain
\begin{equation}
\label{EJ_NESS}
\begin{aligned} 
\lim_{t\to\io}E(x,t)
& = w_{\l} + \sum_{r} \int_{\R^{+}} \frac{dq}{2\pi}
					\frac{\ol(q)}{e^{\b_{r} \ol(q)}-1}, \\
\lim_{t\to\io}J(x,t)
& = \sum_{r} r \int_{\R^{+}} \frac{dq}{2\pi} \frac{d\ol(q)}{dq}
		\frac{\ol(q)}{e^{\b_{r} \ol(q)}-1},
\end{aligned} 
\end{equation}
where $w_{\l}$ is the ground state energy density \cite{MaLi, LLMM}, using that the NESS is translation invariant.
Similarly, for the fermion two-point correlation function, using the well-known bosonization formula expressing fermions as exponentials of plasmons (see, e.g., \cite{HSU, MEJ, LLMM, LaMo} and references therein), we find
\begin{multline}
\label{Sr_NESS}
\lim_{t\to\io} S_{r}(\xi,\tau,x,t)
= \frac{i}{2\pi \ul_{r}}
	\exp
	\biggl(
		\int_{\R^{+}} \frac{dq}{q} \,
		\Bigl\{ e^{iq\ul_{r}(q)} - e^{iq\ul_{r}} \Bigr\}
	\biggr) \\
\begin{aligned}
& \times
	\exp
	\biggl(
		\int_{\R^{+}} \frac{dq}{q} \,
		\sinh^{2} \ph(q) \\
& \qquad \quad
		\times \Bigl\{ e^{iqu_r(q)} +e^{iqu_{-r}(q)} - 2e^{-q0^{+}} \Bigr\}
	\biggr) \\ 
& \times
	\exp
	\biggl(
		\int_{\R^{+}} \frac{dq}{q} \,
		\biggl[
				\cosh^{2} \ph(q) \frac{2\{ \cos[q\ul_{r}(q)]  - 1 \}}{e^{\b_{r}  \ol(q)}-1} \\
& \qquad \quad
			+ \sinh^{2} \ph(q) \frac{2\{ \cos[q\ul_{-r}(q)] - 1 \}}{e^{\b_{-r} \ol(q)}-1}
		\biggr]
	\biggr),
\end{aligned}
\end{multline} 
where $\ul_{r}(p) \equiv r[\x - r\vl(p)\t]+i0^{+}$ and $\ul_{r} \equiv \ul_{r}(0)$.

The second integral in \eqref{EJ_NESS} gives the final energy flow and appears to depend on the interactions.
However, by the change of variables $u = \b_{r} \ol(q)$, we obtain
\begin{equation}
\label{JNESS}
\lim_{t\to\io} J(x,t)
= \sum_{r} r\frac{\pi T_r^2}{12}
= \frac{\pi}{12}(T_L^2-T_R^2)
\equiv J,
\end{equation}
due to the presence of the group velocity $d\ol(q)/dq$ in the integrand [assuming $d\ol(q)/dq > 0$, which is true for a large class of interaction potentials \cite{LLMM}].
It follows that the final heat current only depends on $T_{L,R}$ and is independent of microscopic details.
Such universal behavior, previously observed in CFTs \cite{BeDo1, BeDo2, BeDo3}, thus remains true for the Luttinger model even when scale invariance is broken by the interactions.
This result supports the conjecture, based on numerical simulations of the $XXZ$ chain \cite{KIM}, that for interacting fermions, $J = f(T_L) - f(T_R)$, where, in general, $f$ is a nonuniversal function tending to the universal CFT result \cite{BeDo1} in the low-temperature limit. 

For noninteracting fermions, the temperature dependence $J = f(T_L) - f(T_R)$ corresponds to the above-mentioned factorization of the NESS and was previously obtained analytically by different methods \cite{HoAr, AJPP, BJP, ALA}.
In fact, using these analytical results, the function $f$ for the $XX$ chain can be computed analytically, $f(T) = ({\pi}/{12}) T^2 \left[ 1 - R(b_+) - R(b_-) \right]$, with nonuniversal corrections $R(b_{\pm}) = (6/\pi^2) \int_{b_{\pm}}^{\io} dx\, x/(e^{x}+1)$ governed by $b_{\pm} = ({\vF}/{T a_0}) [{1 \pm \cos(\nu \pi)}]/{\sin(\nu \pi)}$, where $a_0$ is the lattice spacing and $0 < \nu < 1$ is the filling factor (specifying the Fermi momentum $\kF = \nu \pi/a_0$).
If $T a_0 / \vF$ is small, the corrections are exponentially suppressed, and the universal result becomes exact in the {\it scaling limit} $Ta_0/ \vF \to 0$.

The first integral in \eqref{EJ_NESS} expresses the energy density in the NESS as a sum of energy densities at equilibrium with temperatures $T_{L,R}$ and is nonuniversal.
Indeed, it depends on the interactions, and only in the local case, when $\vl(p) = \vl \equiv \vl(0)$ and $\ph(p) = \ph$ are constant, does it simplify to
\begin{equation}
\label{EENESS}
\lim_{t\to\io} E(x,t)
= \sum_r \frac{\p}{12 \vl} T_{r}^{2}
= \frac{\pi}{12\vl}(T_L^2+T_R^2) 
\end{equation}
after an additive renormalization corresponding to subtracting the (diverging) constant $w_{\l}$. 
Similarly, the two-point correlation function in the local case, after a multiplicative renormalization of the fermion fields (not needed in the nonlocal case), becomes
\begin{multline}
\label{Sr_NESS_loc}
\lim_{t\to\io}S_r(\Dx,\Dt,x,t) \\
= \frac{1}{2\pi\tilde{\ell}} \biggl(
		\frac{i \pi T_{r}\tilde{\ell}/\vl}{\sinh(\pi T_{r}\ul_{r}/\vl)}
	\biggr)^{1+\hl/2}
	\biggl(
		\frac{i \pi T_{-r}\tilde{\ell}/\vl}{\sinh(\pi T_{-r}\ul_{-r}/\vl)}
	\biggr)^{\hl/2},
\end{multline}
where $u_{r} = r[\x - r\vl\t]+i0^{+}$, with the equilibrium anomalous exponent $\hl = 2\sinh^{2} \ph$ \cite{MaLi} and a length parameter $\tilde{\ell}$ due to the renormalization; cf.\ also \cite{LLMM, LaMo}.
This exponent depends on the interactions and is nonzero if the interactions are nonzero.
Clearly, unless $\hl=0$, the NESS does \emph{not} factorize into right- (left-) moving fermions with temperatures $T_L$ ($T_R$) as for the $XX$ chain.

%%%%%%%%%%%%%%%%%%%%%%%%%%%%%%%%%%%%%%%%%%%%%%%%%%%%%%%%%%%%%%%%%%%%%%%%%%%%%

\section{Finite-time results: Local interactions}
\label{Sec:FT_results:Local_interactions}
The Luttinger model with local interactions is conformally invariant, implying that $\cH(x,t)$ and $\cJ(x,t)$ satisfy the wave equation, and thus
\begin{equation}
\label{f_res_cHJ_exact}
\begin{aligned}
E(x,t)
& = \frac{1}{2} \left[ G(x-\vl t) + G(x+\vl t) \right], \\
J(x,t)
& = \frac{\vl}{2} \left[ G(x-\vl t) - G(x+\vl t) \right],
\end{aligned} 
\end{equation}
for some function $G(x)$.
Using Method 2 $G(x)$ can be computed as a series expansion in $\eps$ to all orders (see the Appendix), and, after summation, we obtain the following remarkably simple result:
\begin{equation}
\label{zak1}
\begin{aligned}
G(x)
& = \frac{\pi}{6\vl}\frac1{\b(x)^2}
		+ \frac{\vl}{12\pi} \left(
				\frac{\b''(x)}{\b(x)} - \frac{1}{2} \left( \frac{\b'(x)}{\b(x)} \right)^2
			\right) \\
& = \frac{\pi}{6\vl}T(x)^2
		- \frac{\vl}{12\pi} \left(
				\frac{T''(x)}{T(x)} - \frac{3}{2} \left( \frac{T'(x)}{T(x)} \right)^2
			\right).
\end{aligned}
\end{equation}
The term $({\pi}/{6\vl})T(x)^2$ is expected from the equilibrium result for a {\it uniform} temperature profile, but the presence of the derivative terms has apparently been overlooked in the previous literature. 
Thus, in the case of a {\it nonuniform} temperature profile, \eqref{f_res_cHJ_exact} and \eqref{zak1} show that $E(x,t)$ and $J(x,t)$ also depend on the first and second derivatives of $T(x)$, but not on higher-order ones.
This is true even at $t = 0$.

The evolution of the energy flow can be easily understood using \eqref{f_res_cHJ_exact} and \eqref{zak1}.
For a step-like $\b(x) = \b [1+\eps W(x)]$ with $W(x) = - (1/2) \tanh (x/\dw)$, as in the Introduction, the energy profile at $t=0$ away from $x = 0$ is essentially proportional to the local temperature, i.e., $E(x,0)$ equals $({\pi}/{12\vl}) T_{L,R}^2$ far to the left and right.
However, in the transition region, for small $\dw > 0$ and $\eps \neq 0$, the derivative terms in \eqref{zak1} produce peaks; see Fig.~\ref{Fig:ed_with_Tx_insert_Luttinger}. 
As $t$ increases, a region develops around the origin with a uniform energy density bounded by two {\it rigid} fronts (their shape does not change with time) that move ballistically to the right (left) with constant velocity $\vl$ ($-\vl$); see Fig.~\ref{Fig:ed_g_evol_Luttinger}.
In the same region the current has a nonvanishing constant value.
For large times we recover the results for the NESS in \eqref{JNESS} and \eqref{EENESS}.

As we discuss in Sec.~\ref{Sec:FT_results:Nonlocal_interactions}, peaks qualitatively similar to those described above are seen in other related models, including interacting lattice models, such as the $XXZ$ chain, and noninteracting models, such as the $XX$ chain.
It is important to stress that the shape of the peaks is nonuniversal and depends on short-distance details: this is clear already from the interaction dependence of the 
derivative terms that appear in \eqref{f_res_cHJ_exact} due to \eqref{zak1}.

It is interesting to note that $G(x)$ can be written as
\begin{equation}
\label{zak2}
G(x) = \frac{\pi T^2}{6\vl} g'(x)^2 - \frac{\vl}{12\pi} (Sg)(x),
\end{equation}
using the function $g(x) = \int_{0}^{x} dx' \, T(x')/T$ and the so-called Schwarzian derivative \cite{FMS}
\begin{equation}
\label{Schwarzian_derivative}
(Sg)(x) = \frac{g'''(x)}{g'(x)} - \frac{3}{2} \left( \frac{g''(x)}{g'(x)} \right)^2.
\end{equation}
By recalling that the Luttinger model with local interactions is a CFT with central charge $c = 1$, this result has a simple interpretation as follows. 
In a CFT, the energy and heat current densities are given by expectation values of the {\it renormalized energy-momentum tensor} $\mathcal{T}(z)$ and $\bar{\mathcal{T}}(\bar{z})$,
\begin{equation}
\label{f_res_cHJ_exact_CFT}
\begin{aligned}
E(z,\bar{z})
& = - \frac{1}{2\pi} \left[
			\langle \mathcal{T}(z) \rangle + \langle \bar{\mathcal{T}}(\bar{z}) \rangle
		\right], \\
J(z,\bar{z})
& = - \frac{iv}{2\pi} \left[
			\langle \mathcal{T}(z) \rangle - \langle \bar{\mathcal{T}}(\bar{z}) \rangle
		\right], \\
\end{aligned}
\end{equation}
using $z = x + i\vl\t$ and $\bar{z} = x - i\vl\t$, with $\t$ denoting imaginary time \cite{FMS, SoCa}.
Moreover, under a conformal transformation $z \to w(z)$, the renormalized energy-momentum tensor in a CFT transforms as $\mathcal{T}(z) \to \mathcal{T}(w)$, with 
\begin{equation}
\label{EM_tensor_transformation_rule}
\mathcal{T}(z)
= \left( \frac{dw}{dz} \right)^{2} \mathcal{T}(w) + \frac{c\vl}{12} (Sw)(z),
\end{equation}
using the Schwarzian derivative $S$ \cite{FMS}.
From the above one obtains \eqref{f_res_cHJ_exact} by a Wick rotation $\tau \to it$ and the identification $G(x) = -\pi^{-1} \langle \mathcal{T}(z) \rangle|_{z = x}$, using that $E(x,t) = E(z,\bar{z})$ and $J(x,t) = -iJ(z,\bar{z})$. 
Our results in \eqref{f_res_cHJ_exact} and \eqref{zak2} are therefore equivalent to what one would obtain by a conformal transformation determined by the function $g(x) = \int_{0}^{x} dx'\, T(x')/T$ from the equilibrium result $\langle \mathcal{T}(w) \rangle_{\b} = \langle \bar{\mathcal{T}}(\bar{w}) \rangle_{\b} = -c\pi^2 T^2/6\vl$ (for the latter see, e.g., \cite{BeDo3}).
As we discuss in Sec.~\ref{Sec:Conclusions}, it would be interesting to check if this is true also for other observables. 

%%%%%%%%%%%%%%%%%%%%%%%%%%%%%%%%%%%%%%%%%%%%%%%%%%%%%%%%%%%%%%%%%%%%%%%%%%%%%

\section{Finite-time results: Nonlocal interactions}
\label{Sec:FT_results:Nonlocal_interactions} 
We now consider the Luttinger model with nonlocal interactions.
Such interactions break conformal invariance and give rise to dispersion effects since the renormalized Fermi velocity $\vl(p)$ depends on momenta.
These effects are qualitatively similar to ones observed in lattice models.
(The interaction range introduces a scale similar to the lattice spacing.)
We compute quantities only to first order in $\eps$ using Method 1.
Comparison with our all-order results for the NESS and for finite times in the local case suggests that such a first-order approximation works well for small $\eps$: for example, for $\eps = -0.01$, used below in Figs.~\ref{Fig:ed_0_at_t0_and_evolution_Luttinger}~and~\ref{Fig:jd_vs_t_Luttinger}, first- and all-order results are practically indistinguishable, and thus the deviations seen in these figures between the plots for local and nonlocal interactions can be fully attributed to dispersive effects. 

For the energy and heat current densities, we obtain
\begin{equation}
\label{i_res_cHJ}
\begin{aligned}
E(x,t)
& = E_{0} + \eps E_{1}(x,t) + O(\eps^2), \\
J(x,t)
& = \eps J_{1}(x,t) + O(\eps^2),
\end{aligned}
\end{equation}
where $E_{0}$ is equal to $\lim_{t\to\io} E(x,t)$ in \eqref{EJ_NESS} for $\b_{+} = \b_{-} =\b$,
\begin{equation}
\label{kk}
\begin{aligned}
E_1(x,t)
& = - \sum_{r_1,r_2} \strokedint_{\R} \frac{dp}{2\pi} \int_{\R} \frac{dq}{4\pi} \,
			\hat{W}(p) A(p-q,q), \\
J_1(x,t)
& = - \sum_{r_1,r_2} \strokedint_{\R} \frac{dp}{2\pi} \int_{\R} \frac{dq}{4\pi} \,
			\hat{W}(p) \frac{i}{p} \frac{\partial}{\partial t} A(p-q,q),
\end{aligned}
\end{equation}
with
\begin{multline*}
A(p_1,p_2) = e^{i(p_1+p_2)x - i[r_1\ol(p_1)+r_2\ol(p_2)]t} \\
\begin{aligned}
	& \times
		\frac{[r_1\vl(p_1)+r_2\vl(p_2)]^2}{4\vl(p_1)\vl(p_2)}
		\frac{[r_1e^{2\ph(p_1)}+r_2e^{2\ph(p_2)}]^2}{4e^{2[\ph(p_1)+\ph(p_2)]}} \\
	& \times
		\frac{e^{\b[r_1\ol(p_1)+r_2\ol(p_2)]} - 1}{r_1\ol(p_1)+r_2\ol(p_2)}
		\frac{r_1\ol(p_1)}{e^{\b r_1\ol(p_1)}-1}
		\frac{r_2\ol(p_2)}{e^{\b r_2\ol(p_2)}-1}.
\end{aligned}
\end{multline*}
Similarly, for the two-point correlation function, we obtain
\begin{equation}
\label{Sr_Dx_Dt_x_t}
S_{r}(\Dx,\Dt,x,t)
= \langle \psi^+_r(\Dx,\Dt)\psi^-_{r}(0,0) \rangle_{\b}
	e^{\epsilon B_{1;r}(\Dx,\Dt,x,t) + O(\epsilon^2)},
\end{equation}
where $\langle \psi^+_r(\Dx,\Dt)\psi^-_{r}(0,0) \rangle_{\b}$ is equal to $\lim_{t\to\io} S_{r}(\xi,\tau,x,t)$ in \eqref{Sr_NESS} for $\b_{+} = \b_{-} = \b$,
\begin{equation}
\label{i_res_S_1}
B_{1;r}(\Dx,\Dt,x,t)
= - \sum_{r_1, r_2}
		\strokedint_{\R} \frac{dp}{2\pi} \int_{\R} \frac{dq}{4\pi} \,
		\hat{W}(p) C(p-q,q),
\end{equation}
with
\begin{multline*}
C(p_1,p_2)
= 2\pi e^{i(p_1+p_2)x - i[r_1\ol(p_1)+r_2\ol(p_2)]t} \\
\begin{aligned}
	& \times
		\frac{r_1\vl(p_1)+r_2\vl(p_2)}{2}
		\frac{r_1e^{\ph(p_2)-\ph(p_1)}+r_2e^{\ph(p_1)-\ph(p_2)}}{} \\
	& \times
		\frac{e^{\b [r_1\ol(p_1)+r_2\ol(p_2)]} - 1}{r_1\ol(p_1)+r_2\ol(p_2)}
		F_{r}^{r_1}(p_1)
		F_{r}^{r_2}(p_2)
\end{aligned}
\end{multline*}
and
\begin{equation*}
F_{r}^{r'}(p')
= \frac{e^{-\varphi(p')} + rr' e^{\varphi(p')}}{2}
	\frac{e^{i p'r' \ul_{r'}(p')} - 1}{e^{\b r'\ol(p')} - 1}.
\end{equation*}
The above results agree, to first order in $\eps$, with \eqref{EJ_NESS} and \eqref{Sr_NESS}
as $t \to \io$.

Similar to the discussion for the local case in Sec.~\ref{Sec:FT_results:Local_interactions} for a step-like $\b(x)$, our analytical results in \eqref{i_res_cHJ} and \eqref{kk} show, for small $\dw > 0$ and $\eps \neq 0$, that peaks are produced in the transition region between different temperatures; see Fig.~\ref{Fig:ed_with_Tx_insert_Luttinger}.
As $t$ increases, a region develops around the origin with a uniform energy density bounded by two ballistically moving {\it nonrigid} fronts (their shape changes with time); see Fig.~\ref{Fig:ed_g_evol_Luttinger}.

\begin{figure}[!htbp]
\centering

\subfigure[\;At $t=0$]{
\includegraphics[width=0.95\columnwidth, trim=30 0 0 5, clip=true]{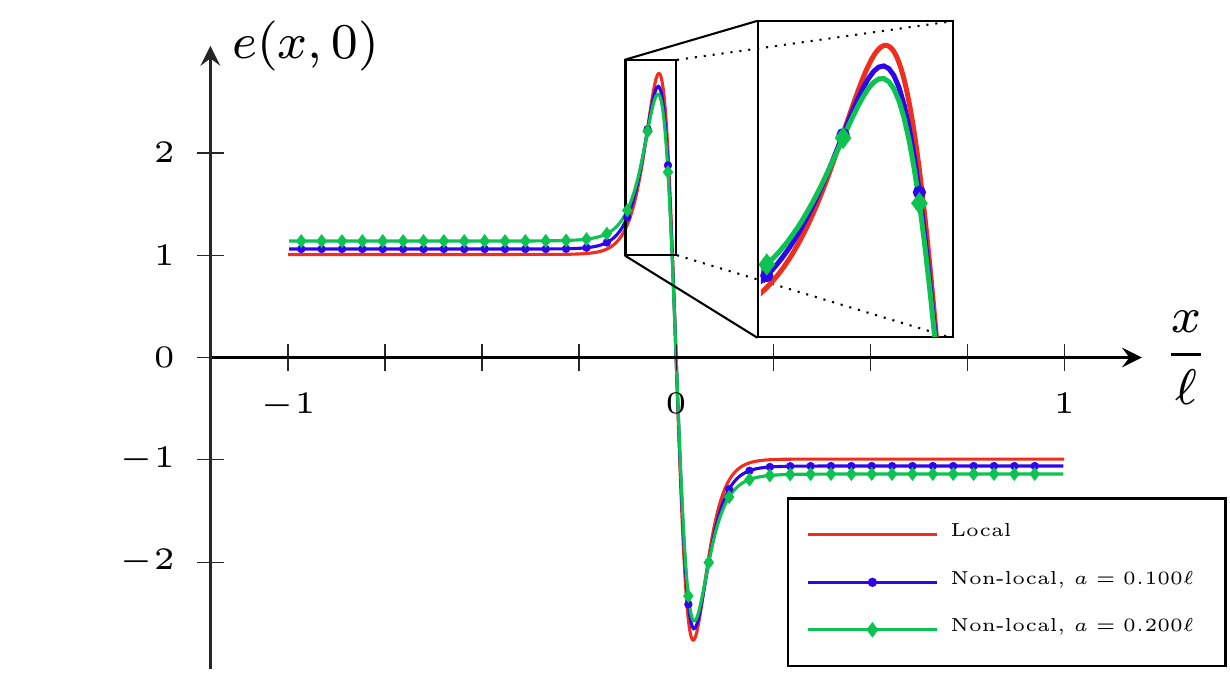}
\label{Fig:ed_with_Tx_insert_Luttinger}
}

\subfigure[\;Evolution for $t > 0$]{
\includegraphics[width=0.95\columnwidth, trim=30 0 0 5, clip=true]{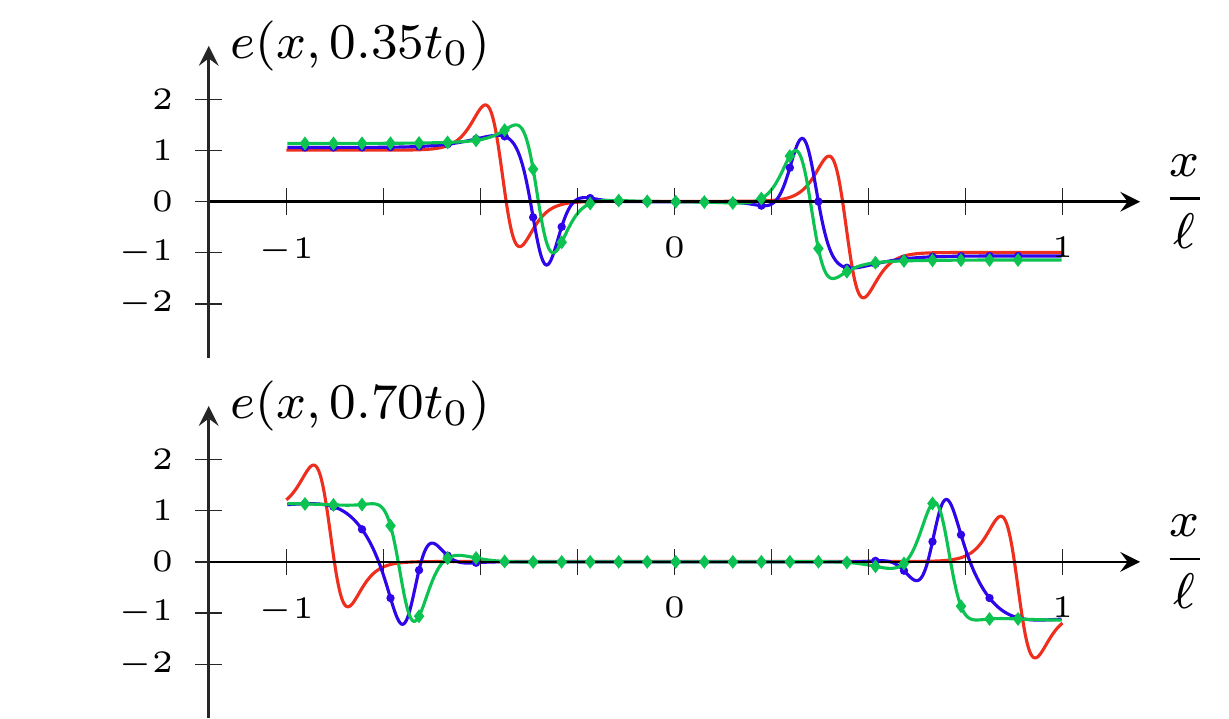}
\label{Fig:ed_g_evol_Luttinger}
}

\caption{Interacting fermions. Plots of analytical results for the energy density $e(x,t) = \vl[E(x,t)-E_{0}]/J$ in an interval $[-\ell,\ell]$ around $x = 0$ at times \subref{Fig:ed_with_Tx_insert_Luttinger} $t = 0$ and \subref{Fig:ed_g_evol_Luttinger} $t > 0$ for the Luttinger model with local and nonlocal interactions.
The results in the local case are given by \eqref{f_res_cHJ_exact} for $V(x) = \pi\vF \d(x)/2$ and in the nonlocal case by \eqref{i_res_cHJ} for $V(x) = \pi\vF/[4a\cosh(\pi x/2a)]$ with $a = 0.100\ell$ and $a = 0.200\ell$, respectively.
The coupling constant is $\l = 0.6$, and the other parameters are $\b = 20$, $\eps = -0.01$, $\dw = 0.06 \ell$, $t_{0} = \ell/\vF$, and $\vF = 1$.
The value of $\eps$ is small enough that $O(\eps^2)$ corrections are negligible.}
\label{Fig:ed_0_at_t0_and_evolution_Luttinger}
\end{figure}
\begin{figure}[!htbp]
\includegraphics[width=0.95\columnwidth, trim=11 3 0 3, clip=true]{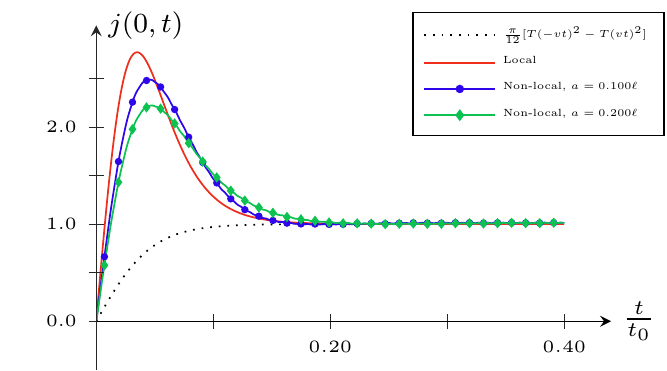}
\caption{Interacting fermions. Plots of analytical results for the heat current $j(0,t) = J(0,t)/J$ through $x=0$ for the Luttinger model using the same parameters as in Fig.~\ref{Fig:ed_0_at_t0_and_evolution_Luttinger}.
Also included is the local case without the second term in \eqref{zak2} (black dotted line).}
\label{Fig:jd_vs_t_Luttinger}
\end{figure}

In Fig.~\ref{Fig:jd_vs_t_Luttinger} we plot the current through $x = 0$ as a function of time.
The plotted results contain an initial peak.
As seen from the dotted line in Fig.~\ref{Fig:jd_vs_t_Luttinger}, such a peak is absent in the local case if the second term in \eqref{zak2} is omitted.
A qualitatively similar peak is present in numerical results for the $XXZ$ chain; see, e.g., Fig.~1(a) in \cite{KIM} and Fig.~3 in \cite{LVMR} showing the heat current through the contact point in the partitioning protocol.
As emphasized in Sec.~\ref{Sec:FT_results:Local_interactions} and also in \cite{KIM}, the shape of such peaks is nonuniversal: in the Luttinger model the shape depends on the interactions and in the $XXZ$ chain on the anisotropy and the dispersion relation.
However, the presence of the peaks seems to be a generic feature.

\begin{figure}[!htbp]
\centering
\subfigure[\;At $t=0$]{
\includegraphics[width=0.95\columnwidth, trim=30 0 0 5, clip=true]{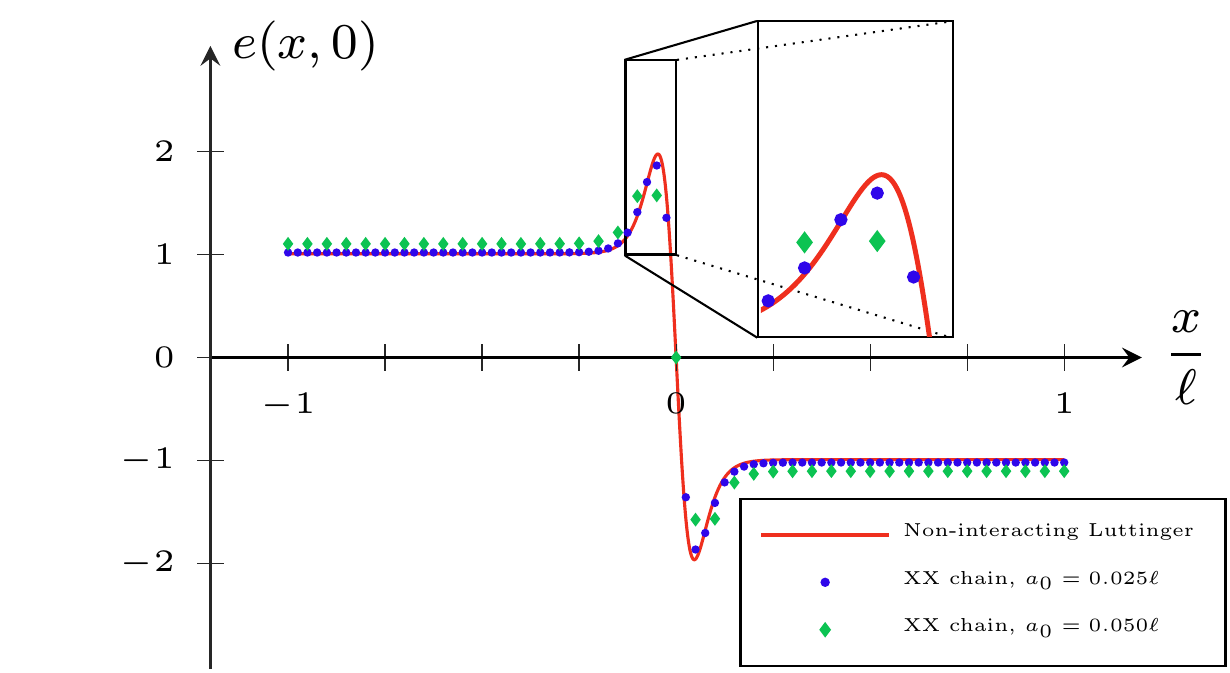}
\label{Fig:ed_with_Tx_insert_XX}
}
\subfigure[\;Evolution for $t > 0$]{
\includegraphics[width=0.95\columnwidth, trim=30 0 0 5, clip=true]{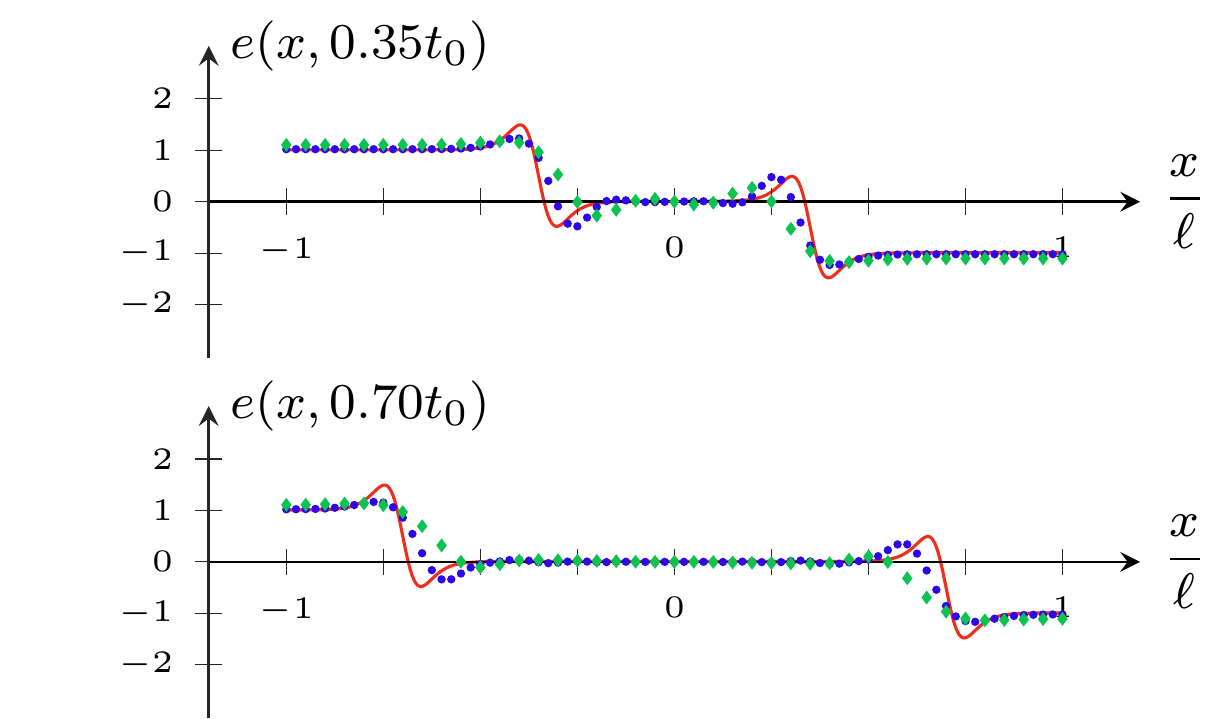}
\label{Fig:ed_g_evol_XX}
}
\caption{Noninteracting fermions. Plots of analytical results for the energy density $e(x,t) = \vl[E(x,t)-E_{0}]/J$ in an interval $[-\ell,\ell]$ around $x = 0$ at times \subref{Fig:ed_with_Tx_insert_XX} $t = 0$ and \subref{Fig:ed_g_evol_XX} $t > 0$ for the noninteracting Luttinger model and for the $XX$ chain.
The results for the former are given by \eqref{f_res_cHJ_exact} and \eqref{zak1} with $v_F$ instead of $v$.
The $XX$ chain is considered close to half filling on a lattice with spacing $a_0 = 0.025 \ell$ and $a_0 = 0.050 \ell$, respectively.
The other parameters are the same as in Fig.~\ref{Fig:ed_0_at_t0_and_evolution_Luttinger}.}
\label{Fig:ed_0_at_t0_and_evolution_XX}
\end{figure}

\begin{figure}[!htbp]
\includegraphics[width=0.95\columnwidth, trim=11 3 0 3, clip=true]{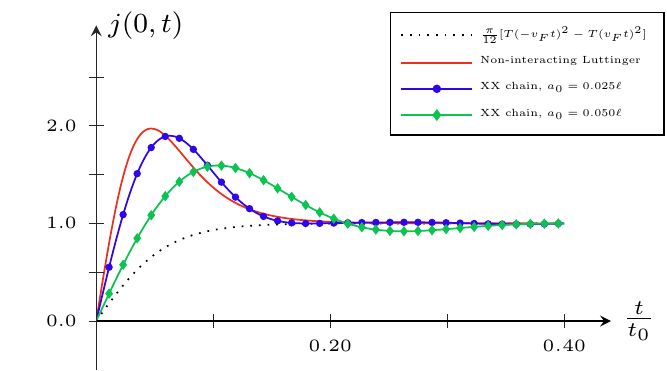}
\caption{Noninteracting fermions. Plots of analytical results for the heat current $j(0,t) = J(0,t)/J$ through $x=0$ for the noninteracting Luttinger model and for the $XX$ chain using the same parameters as in Fig.~\ref{Fig:ed_0_at_t0_and_evolution_XX}.} 
\label{Fig:jd_vs_t_XX}
\end{figure}

To further support our claim about the peaks, we also present, as an example for noninteracting lattice fermions, plots of the corresponding results for the $XX$ chain computed to first order in $\eps$ using Method 1; see Figs.~\ref{Fig:ed_0_at_t0_and_evolution_XX}~and~\ref{Fig:jd_vs_t_XX}.
Peaks and dispersion effects that are qualitatively similar to the ones in Figs.~\ref{Fig:ed_0_at_t0_and_evolution_Luttinger} and \ref{Fig:jd_vs_t_Luttinger} are clearly visible.
Moreover, we checked numerically and analytically, to first order in $\eps$, that the results for the $XX$ chain approach those of the \emph{noninteracting} Luttinger model in the scaling limit; plots of the latter are given by the red (plain) line in Figs.~\ref{Fig:ed_0_at_t0_and_evolution_XX}~and~\ref{Fig:jd_vs_t_XX}.
This is true even at finite times.

%

%%%%%%%%%%%%%%%%%%%%%%%%%%%%%%%%%%%%%%%%%%%%%%%%%%%%%%%%%%%%%%%%%%%%%%%%%%%%%

\section{Methods}
\label{Sec:Methods}
Our results are based on rigorous bosonization methods well known from studies of the Luttinger model in equilibrium; see, e.g., \cite{MaLi, HSU, LLMM, LaMo}.
We work on the circle $-L/2 \leq x \leq L/2$ of length $L > 0$ with the fermion fields $\psi_{r}^{\pm}(x)$ satisfying anti-periodic boundary conditions and take the thermodynamic limit $L\to\io$ only after computing expectation values for finite $t \geq 0$.
The order, first $L \to \io$ and then $t \to \io$, is important for computing results in the long-time limit \cite{LLMM, BeDo1}.

\subsection{Method 1}
\label{SubSec:Method1}
To compute $\langle \cO \rangle$, we write $\cG$ in \eqref{dm_neq} as $\b (H + \cW)$ with $\cW = \eps \int dx \, W(x) \cHl(x)$ and use the fact that $U(\b) \equiv  e^{\b \Hl}e^{-\b(\Hl + \cW)}$ satisfies
\begin{equation}
\partial_{\b} U(\b) = - e^{\b \Hl}\cW e^{-\b(\Hl + \cW)} = - \cW(\b) U(\b),
\end{equation}
with $\cW(\b) = e^{\b \Hl} \cW e^{-\b \Hl}$.
Solving this by iteration we obtain a series expansion in $\eps$ (the Dyson series),
\begin{equation}
\label{DSLR}
\langle \cO \rangle
= \langle \cO \rangle_{\b}
- \eps \left[ \langle \cC\cO \rangle_{\b} 
		- \langle \cC \rangle_{\b} \langle \cO \rangle_{\b} \right]
+ O(\eps^2),
\end{equation}
with $\cC = \int_{0}^{\b} d\b' \int dx \, W(x) \cHl(x,-i\b')$.
It follows that nonequilibrium expectation values are expressed as sums of equilibrium ones.
This method can be used for computing nonequilibrium results to first order in $\eps$ for any model where equilibrium results are computable.
Computations of the energy and heat current densities and the two-point correlation functions for the Luttinger model are straightforward but tedious using Wick's theorem; the details will be presented elsewhere.

\subsection{Method 2}
\label{SubSec:Method2}
Higher-order terms can be computed using general mathematical results for quasi-free models; see, e.g., \cite{GrLa}.
For the bosonized Luttinger Hamiltonian we write $H = d\hat{\G}(K)$ to mean boson second quantization of the one-particle operator $K$, and similarly $\cW = d\hat{\G}(W)$ for some $W$.
(We note that the second quantization map $d\hat{\G}$ is in a nontrivial representation of the boson field algebra and that there are certain technical requirements on the one-particle operators \cite{GrLa, NNS} that are fulfilled in the cases of interest to us.)
For $\cO = d\hat{\G}(O)$ with some one-particle operator $O$, one can show (e.g., using results in \cite{GrLa}) that $\langle \cO \rangle - \langle \cO \rangle_{\b}$ can be written as
\begin{multline}
\label{OPHS}
\frac{\Tr[e^{-\b d\hat{\G}(K+W)} d\hat{\G}(O)]}{\Tr[e^{-\b d\hat{\G}(K+W)}]}
- \frac{\Tr[e^{-\b d\hat{\G}(K)} d\hat{\G}(O)]}{\Tr[e^{-\b d\hat{\G}(K}]} \\
\begin{aligned}
& = \tr \bigl(
		\bigl\{
			[ e^{2\b (K+W)} - 1 ]^{-1} - [ e^{2\b K} - 1 ]^{-1}
		\bigr\} O
	\bigr) \\
& = \frac{1}{\b} \sum_{\nu}
	\tr \bigl(
		\bigl\{ [i\nu - 2(K+W)]^{-1} - [i\nu - 2K]^{-1} \bigr\} O
	\bigr) \\
& = \sum_{n=1}^{\io} \frac{1}{\b}
	\sum_{\nu} \tr \bigl( [i\nu-2K]^{-1} (2W[i\nu-2K]^{-1})^{n} O \bigr),
\end{aligned}
\end{multline}
where $\tr$ is the one-particle trace and the $\nu$ sum is over all boson Matsubara frequencies $\nu \in (2\pi/\b) \mathbb{Z}$.
[Note that the second and third identities in \eqref{OPHS} are standard expansions.]
The computation of $G(x)$ in \eqref{zak1} using \eqref{OPHS} for the Luttinger model with local interactions is explained in the Appendix.

%%%%%%%%%%%%%%%%%%%%%%%%%%%%%%%%%%%%%%%%%%%%%%%%%%%%%%%%%%%%%%%%%%%%%%%%%%%%%

\section{Conclusions}
\label{Sec:Conclusions}
We derived analytical results for the NESS and for the full time evolution of the Luttinger model with both local and nonlocal interactions starting from a nonequilibrium state defined by a smooth nonuniform temperature profile.
These results were computed using methods based on a series expansion in the distance $\eps$ from equilibrium in the initial state.
We showed that the NESS is factorized in terms of the eigenmodes of the interacting Hamiltonian and that its fermion two-point correlation function contains interaction-dependent exponents.
On the contrary, the final heat current is equal to the universal CFT result \cite{BeDo1} even if conformal invariance is broken by the interactions.
Moreover, the form of the temperature dependence of the heat current agrees with the one found numerically in \cite{KIM} for interacting fermions and analytically in \cite{HoAr, AJPP, BJP, ALA} for noninteracting fermions.

For local interactions (and thus {\it a priori} for the noninteracting case), the series for the energy and heat current densities were computed to all orders in $\eps$ and summed into simple exact formulas valid at all times.
These formulas contain a Schwarzian-derivative term [cf.\ \eqref{f_res_cHJ_exact} and \eqref{zak2}], which captures a qualitative feature that appears rather generically, namely, the presence of nonuniversal peaks at short times in the transition region between different temperatures.
We also showed that these formulas coincide with the result obtained by a particular conformal transformation from the corresponding equilibrium result. 
It would be interesting to find an explanation for this and to check if this is true also for other observables and in other CFT models; if true, this would be similar in spirit to results in \cite{SoCa} but for a different physical situation.
Also, it would be interesting to investigate if this can be used to gain some insight into nonequilibrium properties of interacting lattice models, such as the $XXZ$ chain.

For nonlocal interactions, we computed the time evolution of the energy and heat current densities and of the fermion two-point correlation function to first order in $\eps$.
This truncated expansion can be seen as a linear-response approach (cf., e.g., \cite{KIM}) and can, in principle, be used even for models that are not exactly solvable.

%%%%%%%%%%%%%%%%%%%%%%%%%%%%%%%%%%%%%%%%%%%%%%%%%%%%%%%%%%%%%%%%%%%%%%%%%%%%%

\begin{acknowledgments}
We would like to thank Natan Andrei, Jens H.\ Bardarson, Spyros Sotiriadis, and Herbert Spohn for valuable discussions.
E.L.\ acknowledges support by VR Grant No.\ 2016-05167 and thanks the Erwin Schr\"odinger Institute (ESI) in Vienna for their hospitality during the workshop ``Synergies between Mathematical and Computational Approaches to Quantum Many-Body Physics,'' which provided useful input to this work.
The work of J.L.L.\ is supported by AFOSR Grant No.\ FA-9550-16-1-0037 and NSF Grant No.\ DMR1104501.
V.M.\ thanks Rutger's University and the Institute for Advanced Study in Princeton, where parts of this work were done, for their kind hospitality.
P.M. is grateful to the organizers of the 2016 EMS-IAMP Summer School in Mathematical Physics in Rome and thanks L'Universit{\`a} degli Studi di Roma ``La Sapienza'' for support during a visit that was important for the completion of this work.
\end{acknowledgments}

%%%%%%%%%%%%%%%%%%%%%%%%%%%%%%%%%%%%%%%%%%%%%%%%%%%%%%%%%%%%%%%%%%%%%%%%%%%%%

% Appendix
\renewcommand{\theequation}{A\arabic{equation}}
\setcounter{equation}{0}

%%%%%%%%%%%%%%%%%%%%%%%%%%%%%%%%%%%%%%%%%%%%%%%%%%%%%%%%%%%%%%%%%%%%%%%%%%%%%

\section*{Appendix: Computational details}
For local interactions, the one-particle operators $K$ and $W$ in \eqref{OPHS} are given by $K_{r,r'}(p,p') = (rvp/2) \d_{r,r'} \d_{p,p'}$ and $W_{r,r'}(p,p') = \eps [rv\sgn(p) \sqrt{|pp'|}/2L] \d_{r,r'} \hat{W}(p-p')$, respectively.
Since $G(x) = E(x,0)$ [cf.\ \eqref{f_res_cHJ_exact}] it follows that $G(x) = \langle \cH(x) \rangle = \langle d\hat{\G}(O) \rangle$, with $O$ given by $O_{r,r'}(p,p') = [rv\sgn(p) \sqrt{|pp'|}/2L] \d_{r,r'} e^{i(p'-p)x}$.
Using \eqref{OPHS} we obtain $G(x) = \sum_{n=0}^{\io} \eps^{n} G_{n}(x)$, where $G_{0}(x) = \pi/6\vl\b^2$ is the equilibrium result and 
\begin{multline}
\label{Gn_oint}
G_{n}(x)
= \int_{\R^{n+1}} \frac{dp_0 \ldots dp_n}{(2\pi)^{n+1}}\,
	\biggl( \prod_{j=0}^{n-1} \hat{W}(p_j-p_{j+1}) \biggr) \\
\times
	\frac{1}{2} \sum_{r} \frac{1}{\b}\sum_{\nu}
	\biggl( \prod_{j=0}^{n} \frac{r\vl p_{j}}{i\nu-r\vl p_{j}} \biggr)	
	e^{i(p_{0}-p_{n})x},
\end{multline}
for $n = 1,2,\ldots$.
While this formula can be generalized to nonlocal interactions, the local case is special in that it is possible to compute the integrals exactly: changing variables to $q_j = p_{j-1} - p_{j}$ for $j = 1,\ldots,n$ and $p = p_n$, and renaming $\nu \to r\nu$, we can write
\begin{equation}
\label{G_n_I_n_q}
G_{n}(x)
= \frac{v}{4\pi} \int_{\R^{n}} \frac{dq_1 \ldots dq_n}{(2\pi)^{n}}\,
	I_{n}({\bf q}) \biggl( \prod_{j=1}^{n} \hat{W}(q_j) e^{iq_j x} \biggr),
\end{equation}
with
\begin{equation}
\label{I_n_q}
I_{n}({\bf q})
= \frac{2}{v} \int_{\R} dp\,
	\frac{1}{\b} \sum_{\nu}
	\left( \prod_{j = 0}^{n} \frac{v(p+Q_{j})}{i\nu - v(p+Q_{j})} \right),
\end{equation}
where ${\bf q} = (q_1,\ldots,q_n)$ and $Q_{j} = \sum_{k=j+1}^{n} q_{k}$.
The integral in \eqref{I_n_q} can be computed exactly, and, after a lengthy computation, we obtained the following remarkably simple result,
\begin{equation}
\label{I_n_q_exact}
I_{n}({\bf q})
\simeq \frac{(-1)^n}{6}
	\left\{ (n+1) \left( \frac{2\pi}{\b v} \right)^2  + 2 q_{1}^2 + (n-1) q_{1}q_{2} \right\},
\end{equation}
where $\simeq$ is defined by $q_{j}q_{k} \simeq q_{1}^2$ if $j = k$ and $q_{j}q_{k} \simeq q_{1}q_{2}$ if $j \neq k$.
Inserting \eqref{I_n_q_exact} into \eqref{G_n_I_n_q} yields
\begin{gather}
\label{Gn_explicit}
G_{n}(x)
= (-1)^n \biggl( \frac{(n+1)\pi}{6\vl\b^2} W(x)^{n} \\
- \frac{\vl}{12\pi}
	\left[ W''(x)W(x)^{n-1} + \frac{n-1}{2} W'(x)^{2}W(x)^{n-2} \right] \biggr). \nn
\end{gather}
Using this, the series $G(x) = \sum_{n=0}^{\io} \eps^{n} G_{n}(x)$ can be summed, giving the result in \eqref{zak1}.

%%%%%%%%%%%%%%%%%%%%%%%%%%%%%%%%%%%%%%%%%%%%%%%%%%%%%%%%%%%%%%%%%%%%%%%%%%%%%

%%%%%%%%%%%%%%%%%%%%%%%%%%%%%%%%%%%%%%%%%%%%%%%%%%%%%%%%%%%%%%%%%%%%%%%%%%%%%

\end{document}